\documentclass[sigconf, nonacm]{acmart}

\usepackage[most]{tcolorbox}
\usepackage{enumitem}
\usepackage{xspace}

\begin{document}

\title{A Systematic Study of Pseudo-Relevance Feedback with LLMs}

\author{Nour Jedidi}
\affiliation{
    \institution{University of Waterloo}
    \city{Waterloo}
    \state{ON}
    \country{Canada}}

\email{njedidi@uwaterloo.ca}

\author{Jimmy Lin}
\affiliation{
    \institution{University of Waterloo}
    \city{Waterloo}
    \state{ON}
    \country{Canada}
}
\email{jimmylin@uwaterloo.ca}

\newcommand{\umbrela}{PRF-\textsc{Umbrela}\xspace}   
\newcommand{\hyde}{HyDE\xspace}  
\newcommand{\hydeprf}{PRF-HyDE\xspace} 
\newcommand{\hybridjudge}{\textsc{Umbrela}-HyDE\xspace}

\begin{abstract}
Pseudo-relevance feedback (PRF) methods built on large language models (LLMs) can be organized along two key design dimensions:  the \emph{feedback source}, which is where the feedback text is derived from and the \emph{feedback model}, which is how  the given feedback text is used to refine the query representation. However, the independent role that each dimension plays is unclear, as both are often entangled in empirical evaluations.  In this paper, we address this gap by systematically studying how the choice of feedback source and feedback model impact PRF effectiveness through controlled experimentation. Across 13 low-resource BEIR tasks with five LLM PRF methods, our results show: (1) the choice of feedback model can play a critical role in PRF effectiveness; (2) feedback derived solely from LLM-generated text provides the most cost-effective solution; and (3) feedback derived from the corpus is most beneficial when utilizing candidate documents from a strong first-stage retriever.  Together, our findings provide a better understanding of which elements in the PRF design space are most important.\footnote{Code is available at \url{https://github.com/castorini/llm-relevance-feedback}} 
\end{abstract}

\maketitle
\pagestyle{empty}
\begin{figure}[t]
\centering
\includegraphics[width=0.75\linewidth]{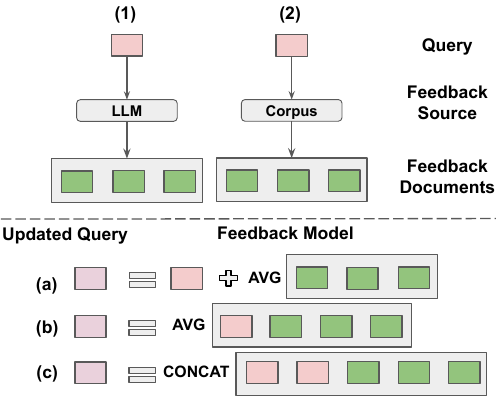}
\caption{Demonstration of feedback source and feedback model. Feedback can come from (1) an LLM, (2) the corpus, or both. Feedback documents then get passed to a feedback model to  update the query. (a) Weighted average vector, (b) average vector, and (c) string concatenation with a query repeat are different feedback models used in the literature.} 
\label{fig:feedback_source_model}
\end{figure}

\begin{figure*}[t!]
  \centering
  \includegraphics[width=0.9\textwidth]{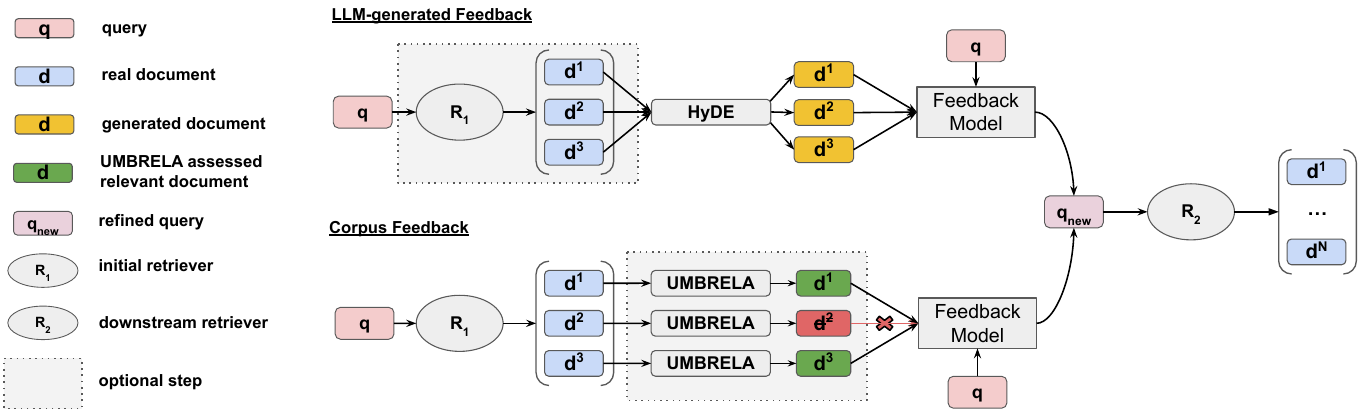}
  \caption{Overview of different PRF pipelines. Dotted boxes denote \emph{optional} steps. For example, if \emph{not} passing blue documents to UMBRELA, $d^2$ gets fed into the feedback model. Running both PRF pipelines in parallel is equivalent to \hybridjudge which concatenates both sets of feedback documents (yellow and green) before passing them to the feedback model.}
  \label{fig:methods}
\end{figure*}

\section{Introduction}

Modern pseudo-relevance feedback (PRF) techniques have increasingly begun to leverage large language models (LLMs) as a tool for improving query representations. The use of LLMs for pseudo-relevance feedback has provided notable improvements in effectiveness compared to traditional approaches which refine query representations using relevance signals from top-ranked documents. 

Current methods that leverage LLMs for pseudo-relevance feedback can be organized along two key design dimensions: the \emph{feedback source} and the \emph{feedback model}. The feedback source refers to where the feedback  -- i.e., the terms or vectors used to improve the query -- comes from, while the feedback model is how this feedback is used to update the query representation. LLM PRF methods typically derive feedback from one of three sources: corpus documents (corpus only), LLM-generated text (LLM only), or a combination (corpus \& LLM). To update the query representation, the approach differs based on the downstream retriever. With BM25, various approaches have been proposed, such as updating via a simple string concatenation~\cite{wang2023query2doc}, adaptively reweighting the query weight based on the length of the feedback~\cite{zhang-etal-2024-exploring-best}, or using classical methods~\cite{jedidi2025revisiting}.  With dense retrievers, LLM PRF methods have primarily utilized a simple average vector update ~\cite{gao2023precise, jedidi2024zero}, but other approaches like Rocchio~\cite{rocchio1971relevance} can be utilized~\cite{li2022pseudo}. 

However, the independent role that the feedback source and feedback model play in the downstream effectiveness of PRF with LLMs remains unclear. Existing LLM PRF approaches often entangle both dimensions in empirical evaluations: proposed methods commit to a specific feedback source and feedback model, but compare to baselines which differ along both dimensions. Not only that, but other factors, such as the number of feedback terms, number of feedback documents, or even downstream retriever are not controlled for. Motivated by this, our goal in this paper is to \emph{fairly} and  \emph{systematically} evaluate these design dimensions. To do so, we organize our paper around the following research questions: 

\begin{itemize}
    \item[{\bf RQ1}]   For a given feedback source, how does the choice of feedback model impact PRF effectiveness? 
    \item[{\bf RQ2}]  Should feedback be derived from high-quality corpus documents or ``fake'' LLM-generated documents? 
    \item[{\bf RQ3}] Does combining feedback sources improve PRF effectiveness over using a single source? If so, how should the feedback be combined?
  \item[{\bf RQ4}]  How does the latency vary among LLM PRF methods?
\end{itemize}

\noindent
To answer these research questions, we reimplement five LLM PRF methods which represent each of the three primary feedback sources (corpus only, LLM only, and corpus \& LLM), ensuring, to the best of our ability, that any confounding factors -- like the number of feedback terms or number of feedback documents -- are controlled for across all methods. We then evaluate each method on 13 low-resource BEIR datasets~\cite{thakur2021beir}, with three different retrievers -- BM25, Contriever, and Contriever
MS-MARCO -- and various feedback models. To support reproducibility, our work directly builds on existing PRF infrastructure present in Anserini~\cite{yang2017anserini} and Pyserini~\cite{lin2021pyserini, li2022pseudo}, enabling our implementations to be seamlessly integrated with widely used open-source IR research toolkits. Our findings can be summarized as follows:

\begin{itemize}[leftmargin=15pt]
    \item For \textbf{RQ1}, we find that the feedback model is an important design choice for methods which derive feedback from LLM-generated text. 
    \item For \textbf{RQ2}, our results demonstrate that feedback should generally be derived from “fake” LLM-generated documents  unless the corpus feedback comes from a strong initial retriever \emph{and} BM25 is utilized as the  downstream retriever.
    \item For \textbf{RQ3}, with the Contriever models, the results show a clear advantage when combining feedback sources that are derived independently and concatenated. For BM25, combining feedback \emph{can} be helpful, but its benefit depends on deriving feedback terms from high-quality corpus documents; beyond that, \emph{how} the feedback is combined does not matter. 
    \item For \textbf{RQ4}, we found LLM only feedback to be the most efficient across the board. When integrating corpus feedback, the latency is dependent on the number of  candidate corpus documents or the average document length in the corpus. 
\end{itemize}

\noindent
Furthermore, our evaluations reveal additional insights, including: 
\begin{itemize}[leftmargin=15pt]
    \item Utilizing BM25 with LLM only feedback is a surprisingly tough baseline to beat. Corpus feedback provides meaningful gains if the LLM integrates more candidate documents, but this comes at the cost of latency. One can also improve effectiveness by integrating candidate documents from a strong initial retriever, but this reduces the relative benefit of PRF. 
    \item Feedback terms derived from high-quality corpus documents retrieved by HyDE are interestingly more effective than the feedback terms generated directly by HyDE, boosting effectiveness by 3.6\%.
    \item Dense retrievers do not fully take advantage of PRF. For example, when provided access to the \emph{same exact} feedback content, PRF with BM25 consistently outperforms PRF with Contriever MS-MARCO by as much as 8.5\%, despite BM25 starting off 10\% less effective than Contriever MS-MARCO. 
\end{itemize}

\noindent
The core contribution of this work is a large-scale systematic study that evaluates different LLM PRF methods ``under the same roof'', in which the \emph{individual} impact of the feedback source and feedback model can be properly understood. By doing so, our results enable a better understanding of which elements in the PRF design space are most critical. We hope our study can provide clarity and guide best practices for future PRF methods built with LLMs.


\section{Background and Related Work}

The traditional setup for pseudo-relevance feedback is as follows: given a query, $q$, an initial retriever, $R_1$, produces a ranked list of documents, $D$. From $D$, the top-$k$ candidate documents,  $\{d_{1}, d_{2}, \dots, d_{k}\}$, are taken to be relevant, forming the set of pseudo-relevant
\emph{feedback documents}, $f_{\text{PRF}}$.   Next, $f_{\text{PRF}}$ gets passed to a \emph{feedback model} which generates the updated query, $q_{new}$. The specifics of how the feedback model generates $q_{new}$ is discussed in Section~\ref{sec: feedback_models}.

With the advent of LLMs, research has recently shifted to  directly utilizing knowledge contained within these models to assist in updating the query representation. With LLMs, the core PRF concept remains: there exists a set of feedback text from which useful terms (or vectors, in the case of dense retrieval) are extracted and utilized for enriching the original query representation. The only difference lies in where the feedback text is derived from. 

Early work focused on training LLMs to generate the feedback text (this is commonly referred to as LLM query expansion in the literature). One of the initial methods in this space was GAR~\cite{mao-etal-2021-generation}, which trained an LLM to expand queries by generating various useful expansion texts given the query as input.  As LLMs started  demonstrating strong \emph{few-shot} and \emph{zero-shot} capabilities~\cite{brown2020language}, methods began generating expansions via direct prompting. Approaches like HyDE~\cite{gao2023precise} and Query2Doc~\cite{wang2023query2doc} focused on hypothetical answer document generation, prompting the LLM to generate text that ``looks like'' a real  document. Researchers also looked into various other prompting strategies~\cite{jagerman2023query} and other forms of generated expansions (e.g., keywords, summaries, or entities)~\cite{mackie2023generative, dhole2024genqrensemble}. 

Follow up methods then began incorporating corpus evidence in combination with the LLM-generated expansions, either by using pseudo-relevant documents as in-context examples for hypothetical answer document generation~\cite{shen-etal-2024-retrieval, jagerman2023query} or fusing representations independently derived from the LLM and the corpus ~\cite{lei-etal-2024-corpus, jia-etal-2024-mill}. Building on this, ThinkQE~\cite{lei-etal-2025-thinkqe}  proposed an approach in which the LLM iteratively refines the query representation through multiple interactions with the corpus. In parallel, other works like ReDE-RF~\cite{jedidi2024zero} and LLMF~\cite{otero2026llm} instead leveraged LLMs as a tool for \emph{selecting}, rather than generating, top-ranked documents from $R_1$ for enriching the query representation. Recently ~\citet{bigdeli2025querygym} released QueryGym, which  standardizes the implementation of these various methods into a singular Python toolkit. 

While the core difference between traditional PRF and LLM PRF lies in where feedback text comes from, how the feedback is utilized for updating the query representation has differed, particularly with sparse retrievers like BM25~\cite{jedidi2025revisiting}. Initially, LLM PRF methods for BM25 simply  generated $q_{new}$ via a simple string concatenation, for example by repeating the query a fixed number of times and appending it to the generated expansion~\cite{wang2023query2doc, jagerman2023query, lei-etal-2024-corpus,shen-etal-2024-retrieval}. MuGI~\cite{zhang-etal-2024-exploring-best} then observed the importance of a proper amount of query repetitions with respect to the expansion content and proposed an adaptive query reweighting strategy that repeats the query a variable number of times based on the length of the generated expansions. 
 Most recently, ~\citet{jedidi2025revisiting} demonstrated that LLM query expansion methods should instead return to classic feedback models like RM3~\cite{abdul2004umass} and Rocchio~\cite{rocchio1971relevance} for updating the query representation, demonstrating strong improvements compared to recent string concatenation approaches, including MuGI. In this paper, for updating the query for BM25,  we focus primarily on classic feedback models -- and not string concatenation methods -- given they are well-established and have demonstrated strong effectiveness with LLM PRF methods. 
 
 On the other hand, with dense retrievers, there has been more consistency in how queries are updated. Methods like HyDE and ReDE-RF only considered a simple average vector update. In this paper, we extend our evaluation to the Rocchio dense vector update, as was done in ~\citet{li2022pseudo}.

\section{Pseudo-Relevance Feedback with LLMs}
\label{sec:methods}

In this section, we overview each of the different approaches for generating -- or selecting -- feedback documents (i.e., the feedback sources) with LLMs. We then discuss the feedback models that are considered for updating the query representation in our evaluations. An overview of the different PRF pipelines is shown in Figure~\ref{fig:methods}. 

\subsection{Feedback Source}

\subsubsection{Corpus Only} To leverage an LLM for selecting feedback documents directly from the corpus, we build upon the approach utilized by ReDE-RF~\cite{jedidi2024zero}. In particular, given $D$, the ranked list of documents retrieved by the initial retriever, we utilize  \textsc{Umbrela}~\cite{upadhyay2024umbrela} to generate a relevance label for each document in the top-$k$ candidate documents, i.e., $\{d_{1}, d_{2}, \dots, d_{k}\}$.
\textsc{Umbrela} then returns a relevance assessment for each document, producing a score between 0-3, where 0 denotes ``irrelevant'', 1 denotes ``related'', 2 denotes ``highly relevant'', and 3 denotes``perfectly relevant''. The output of this step is a list, $\{r(d_{1}), r(d_{2}), \dots, r(d_{k})\}$, where $r(d_{i})$ denotes the score produced by \textsc{Umbrela} for document $d_i$.

Next, using some heuristic, documents deemed relevant are selected for the feedback set, $f_{\text{UMBRELA}}$. In our case, for simplicity, and to avoid the need for hyperparameter tuning, we opt for a simple heuristic for selecting which documents to include in $f_{\text{UMBRELA}}$: for $\{r(d_{1}), r(d_{2}), \dots, r(d_{k})\}$, for a given $q$, we select all documents with the maximum score amongst all the assessed documents, i.e., 
\begin{equation}
    f_{\text{UMBRELA}} = \{ d_i \in f \mid r(d_i) = \max_{d_j \in \{d_{1}, d_{2}, \dots, d_{k}\}} r(d_j) \}.
\end{equation}

\noindent
We refer to this method as \umbrela.

\subsubsection{LLM Only}

For \emph{generating} feedback based on the LLM's parametric knowledge, we leverage HyDE~\cite{gao2023precise}. HyDE samples $N$ hypothetical answer documents by  zero-shot prompting an LLM to write a passage that answers the question. The hypothetical documents, $\hat{d}_i$, generated by the LLM form the feedback set: $f_{\text{HyDE}}=\{\hat{d}_1, \dots, \hat{d}_N\}$. 

\subsubsection{Corpus \& LLM}
\label{sec: llm_corpus_methods}
Finally, we consider two methods which combine feedback derived from the corpus and the LLM's parametric knowledge. As one of our goals is to understand the improvement when combining feedback sources versus using a singular feedback source, we directly build on \umbrela and \hydeprf so we can evaluate the \emph{relative} boost of augmenting a given feedback source with the other feedback source (i.e., augment corpus only feedback with feedback generated by an LLM). 

The first method is motivated by the approach used in CSQE~\cite{lei-etal-2024-corpus}. We independently run both \umbrela and HyDE and then simply concatenate their feedback sets together, i.e.,  $f_{\text{\hybridjudge}} = f_{\text{UMBRELA}} \cup f_{\text{HyDE}}$. We refer to this method as \hybridjudge.

The second method, which we refer to as \hydeprf, simply extends on the HyDE setup by additionally prompting it using the top-$k$ candidate documents, $\{d_{1}, d_{2}, \dots, d_{k}\}$, as \emph{in-context} examples, i.e., ``write a passage to answer the question, using the provided reference passages as context''. In this case the feedback documents, $f_{\text{\hydeprf}}$,   are of the same form as those of HyDE, but instead now HyDE has access to the language, structure, and facts present in the corpus.

Both \hybridjudge and \hydeprf  represent different ways to combine corpus information with the LLM's parametric knowledge. On one hand, \hybridjudge does so \emph{explicitly}: corpus vectors are directly integrated into the feedback model which updates the query vector.   On the other hand, \hydeprf does so \emph{implicitly}: the LLM is still responsible for the content it generates but can be guided by the in-context corpus examples.

\subsection{Feedback Models}
\label{sec: feedback_models}
Given the different sets of feedback -- i.e.,  $f_{\text{UMBRELA}}$, $f_{\text{HyDE}}$,  $f_{\text{\hydeprf}}$, and $f_{\text{\hybridjudge}}$ --  how does the query get updated?  Before describing the feedback models, for consistency in terminology across the different feedback sets, let $f$ be the set of feedback documents that will be used for generating the updated query, $q_{new}$. Let $q_\text{vec}$ be the query vector and $d_\text{vec}[i]$ be the document vector of the $i_{\text{th}}$ document in $f$.  There are differences in the query update for sparse retrievers and dense retrievers, which are described below.

\subsubsection{Sparse retrieval} For updating  $q$ for BM25,  we consider two commonly utilized feedback models in the literature.
The first is the Rocchio algorithm~\cite{rocchio1971relevance}. The Rocchio algorithm computes the weight of a term $t$ in $q_{new}$, $w_{t, q_{\text{new}}}$, using the following equation:

\begin{equation}
    w_{t, q_{\text{new}}}
    = \alpha q_\text{vec}[t] + \frac{\beta}{N} \sum_{d_\text{vec} \in f}  d_\text{vec}[i][t]
    \label{eq:rocchio}
\end{equation}

\noindent
where $q_\text{vec}[t]$ is the frequency of $t$ in $q_\text{vec}$ and  $d_\text{vec}[i][t]$ is the frequency of  $t$ in $d_\text{vec}[i]$. $\alpha$ and $\beta$ are hyperparameters which control the influence of the query terms and feedback document terms. 

The next feedback model we consider is RM3~\cite{abdul2004umass}. We repeat the equation below for reference, but refer the curious reader to ~\citet{abdul2004umass} for specific 
details on how certain components are computed in practice. RM3 computes  $w_{t, q_{\text{new}}}$ using the following equation: 

\begin{equation}
    w_{t, q_{\text{new}}}
    = \lambda P(t|q_\text{vec}) + (1 - \lambda) \sum_{d_\text{vec} \in f} P(t|d_\text{vec}[i])
    \label{eq:rm3}
\end{equation}

\noindent
where $\lambda$ controls how much weight is given to terms in the query versus terms in the feedback documents.  

We note that each of the feedback documents typically undergoes some sort of preprocessing which first removes the common corpus terms and then further prunes the text to a set of candidate expansion terms based on a term’s aggregated importance (e.g., term frequency) across all documents in $f$. For this, we directly follow the implementation procedure in  Anserini~\cite{yang2017anserini}, which is described in more detail by ~\citet{jedidi2025revisiting}. Once each of the term weights $w_{t, q_{\text{new}}}$ is computed, custom Lucene queries are constructed in Pyserini~\cite{lin2021pyserini}, assigning each term a boost equal to its respective weight. The Lucene queries constitute $q_{new}$, which gets passed to the downstream retriever, $R_2$,  for a second round of search. 

\subsubsection{Dense retrieval} For dense retrieval, we consider both average vector feedback and Rocchio vector feedback~\cite{li2022pseudo}. The average vector simply concatenates the vector, $q_\text{vec}$, with each of $d_\text{vec} \in f$ and takes a simple average:

\begin{equation}
     q_{new} = \textrm{Avg}(q_\text{vec}, d_\text{vec}[1], \cdots, d_\text{vec}[N])
    \label{avg}
\end{equation}

\noindent
The Rocchio vector update mimics Equation~\ref{eq:rocchio}:

\begin{equation}
     q_{new} = \alpha q_\text{vec} +  \beta \textrm{Avg}(d_\text{vec}[1], \cdots, d_\text{vec}[N])
    \label{rocchio_dense}
\end{equation}
\noindent
As before, $q_{new}$ then gets passed to $R_2$ for a second round of search.
\begin{table*}[t!]
\centering
\caption{Main results (nDCG@20) on BEIR across LLM PRF methods using \textbf{BM25} as $R_1$ and $R_2$. All LLM-based methods leverage Qwen3-14B as the backbone model. Roc. denotes the Rocchio feedback model.}
\begin{tabular}{l| ccc | cc | cc | cc | cc}
\toprule
\multicolumn{1}{l}{} & \multicolumn{5}{c}{Source: Corpus Only} & \multicolumn{2}{c}{Source: LLM Only} & \multicolumn{4}{c}{Source: Corpus \& LLM} \\
\cmidrule(lr){2-6} \cmidrule(lr){7-8} \cmidrule(lr){9-12}
\multicolumn{1}{l}{} 
& \multicolumn{3}{c}{BM25} 
& \multicolumn{2}{c}{\umbrela} 
& \multicolumn{2}{c}{HyDE} 
& \multicolumn{2}{c}{\hydeprf} 
& \multicolumn{2}{c}{\hybridjudge} \\
\cmidrule(lr){2-4} \cmidrule(lr){5-6} \cmidrule(lr){7-8} \cmidrule(lr){9-10} \cmidrule(lr){11-12}
Dataset & Base & RM3 & Roc. & RM3 & Roc. & RM3 & Roc. & RM3 & Roc. & RM3 & Roc. \\
\midrule
Covid       &  54.6 &	60.3 &	58.7 &	65.8 &	64.9 &	69.6 &	70.7 &	69.3 &	71.5 &	68.1 &	69.4 \\
News          & 37.6	& 44.0 &	43.4 &	48.6 &	48.9 &	47.7 &	49.2 &	49.2 &	50.4 &	49.2 &	50.5 \\
SciFact       & 69.2 &	68.4 &	65.1 &	74.8 &	74.5 &	73.5 &	74.2 &	74.3 &	75.3 &	73.8 &	75.8  \\
FiQA-2018     & 26.0 &	24.5 &	21.8 &	30.3 &	29.4 &	29.2 &	29.2 &	30.5 &	30.5 &	30.2 &	30.3  \\
DBPedia       &  32.4 &	32.5 &	31.9 &	36.6 &	36.4 &	38.7 &	40.8 &	37.1 &	38.9 &	38.6 &	41.1 \\
NFCorpus      & 29.4 &	32.8 &	32.6 &	34.3 &	35.1 &	33.9 &	34.8 &	34.3 &	35.6 &	33.8 &	35.1 \\
Robust04      & 37.6 &	39.9 &	39.3 &	47.9 &	47.8 &	47.6 &	48.6 &	47.6 &	48.6 &	48.8 &	50.2  \\
SCIDOCS      & 17.0 &	17.6 &	17.0 &	19.3 &	19.2 &	18.2 &	18.7 &	18.1 &	18.3 &	19.0 &	19.5 \\
ArguAna     &  33.8 &	43.2 &	39.7 &	48.8 &	46.9 &	45.4 &	45.8 &	45.7 &	46.2 &	49.8 &	47.3  \\
NQ            & 33.2 &	34.4 &	33.1 &	45.0 &	45.0 &	46.1 &	48.9 &	44.5 &	46.6 &	47.3 &	50.1  \\
BioASQ        & 54.9 &	53.5 &	50.8 &	56.7 &	55.5 &	57.7 &	56.8 &	59.2 &	58.7 &	57.9 &	57.4  \\
Climate-FEVER & 18.7 &	21.9 &	22.4 &	23.5 &	23.8 &	28.3 &	29.9 &	26.9 &	27.9 &	27.7 &	28.6  \\
Signal-1M     &  30.3 &	29.8 &	29.6 &	31.2 &	31.3 &	31.3 &	31.3 &	30.8 &	31.2 &	31.5 &	31.8 \\
\midrule
Avg           & 36.5 &	38.7 &	37.3 &	43.3 &	43.0 &	43.6 &	44.5 &	43.7 &	44.6 &	44.3 &	45.2 \\

\bottomrule
\end{tabular}
\label{tab:beir_unsupervised_bm25_only}
\end{table*}

\begin{table*}[t!]
\centering
\caption{Main results (nDCG@20) on BEIR across LLM PRF methods using \textbf{Contriever} as $R_1$ and $R_2$. All LLM-based methods leverage Qwen3-14B as the backbone model. Roc. denotes the Rocchio feedback model.}
\begin{tabular}{l| ccc | cc | cc | cc | cc}
\toprule
\multicolumn{1}{l}{} & \multicolumn{5}{c}{Source: Corpus Only} & \multicolumn{2}{c}{Source: LLM Only} & \multicolumn{4}{c}{Source: Corpus \& LLM} \\
\cmidrule(lr){2-6} \cmidrule(lr){7-8} \cmidrule(lr){9-12}
\multicolumn{1}{l}{} 
& \multicolumn{3}{c}{Contriever} 
& \multicolumn{2}{c}{\umbrela} 
& \multicolumn{2}{c}{HyDE} 
& \multicolumn{2}{c}{\hydeprf}
& \multicolumn{2}{c}{\hybridjudge} \\
\cmidrule(lr){2-4} \cmidrule(lr){5-6} \cmidrule(lr){7-8} \cmidrule(lr){9-10} \cmidrule(lr){11-12}
Dataset & Base & Avg & Roc. & Avg & Roc. & Avg & Roc. & Avg & Roc. & Avg & Roc. \\
\midrule
Covid         & 23.7 & 23.6 & 24.1 & 33.2 & 32.9 & 50.5 & 56.2 & 48.0 & 54.1 & 53.9 & 56.4 \\
News          & 34.0 & 36.6 & 38.2 & 43.5 & 44.2 & 38.0 & 41.9 & 41.4 & 43.3 & 44.5 & 47.1 \\
SciFact       & 66.0 & 49.1 & 58.5 & 72.2 & 71.6 & 64.8 & 69.3 & 70.7 & 70.9 & 71.5 & 71.4 \\
FiQA-2018     & 26.5 & 21.3 & 23.8 & 30.6 & 30.9 & 27.3 & 31.9 & 25.8 & 29.8 & 32.0 & 34.3 \\
DBPedia       & 29.8 & 25.2 & 27.4 & 34.6 & 34.1 & 35.0 & 36.9 & 34.3 & 35.6 & 38.1 & 38.4 \\
NFCorpus      & 29.4 & 25.0 & 27.5 & 31.8 & 32.1 & 23.3 & 27.4 & 29.8 & 31.1 & 30.7 & 31.8 \\
Robust04      & 29.3 & 29.3 & 30.5 & 38.0 & 37.9 & 31.7 & 36.8 & 35.2 & 38.5 & 39.7 & 41.3 \\
SCIDOCS       & 17.4 & 14.2 & 15.5 & 18.6 & 18.7 & 14.4 & 16.7 & 12.3 & 14.8 & 17.6 & 18.6 \\
ArguAna       & 40.9 & 31.0 & 34.9 & 41.9 & 42.9 & 35.0 & 41.4 & 28.1 & 35.7 & 42.8 & 43.8 \\
NQ            & 28.5 & 24.4 & 26.2 & 38.9 & 38.9 & 35.0 & 39.6 & 33.9 & 37.1 & 40.7 & 42.3 \\
BioASQ        & 31.9 & 21.4 & 24.4 & 36.4 & 36.7 & 30.0 & 37.0 & 36.5 & 39.3 & 36.8 & 39.7 \\
Climate-FEVER & 17.6 & 19.0 & 19.5 & 20.4 & 20.2 & 26.7 & 28.1 & 25.4 & 26.2 & 26.3 & 26.8 \\
Signal-1M     & 20.8 & 19.1 & 20.6 & 21.5 & 21.4 & 13.7 & 18.9 & 14.8 & 19.5 & 19.9 & 21.4 \\
\midrule
Avg           & 30.4 & 26.1 & 28.5 & 35.5 & 35.6 & 32.7 & 37.1 & 33.5 & 36.6 & 38.0 & 39.5 \\

\bottomrule
\end{tabular}
\label{tab:beir_unsupervised_contriever}
\end{table*}

\begin{table*}[t!]
\centering
\caption{Main results (nDCG@20) on BEIR across LLM PRF methods using \textbf{Contriever MSM} as $R_1$ and $R_2$. All LLM-based methods leverage Qwen3-14B as the backbone model. Roc. denotes the Rocchio feedback model.}
\begin{tabular}{l| ccc | cc | cc | cc | cc}
\toprule
\multicolumn{1}{l}{} & \multicolumn{5}{c}{Source: Corpus Only} & \multicolumn{2}{c}{Source: LLM Only} & \multicolumn{4}{c}{Source: Corpus \& LLM} \\
\cmidrule(lr){2-6} \cmidrule(lr){7-8} \cmidrule(lr){9-12}
\multicolumn{1}{l}{} 
& \multicolumn{3}{c}{Contriever MSM} 
& \multicolumn{2}{c}{\umbrela} 
& \multicolumn{2}{c}{HyDE} 
& \multicolumn{2}{c}{\hydeprf}
& \multicolumn{2}{c}{\hybridjudge} \\
\cmidrule(lr){2-4} \cmidrule(lr){5-6} \cmidrule(lr){7-8} \cmidrule(lr){9-10} \cmidrule(lr){11-12}
Dataset & Base & Avg & Roc. & Avg & Roc. & Avg & Roc. & Avg & Roc. & Avg & Roc. \\
\midrule
Covid         & 53.4 & 52.1 & 53.9 & 55.4 & 55.3 & 64.0 & 71.1 & 66.3 & 71.3 & 66.8 & 70.5 \\
News          & 41.3 & 38.8 & 40.6 & 44.7 & 45.5 & 44.0 & 45.2 & 46.6 & 47.2 & 45.6 & 46.2 \\
SciFact       & 69.0 & 48.2 & 60.0 & 73.9 & 73.5 & 71.4 & 72.1 & 73.5 & 73.8 & 73.7 & 74.2 \\
FiQA-2018     & 35.5 & 22.8 & 28.7 & 34.7 & 36.1 & 29.8 & 34.3 & 30.5 & 34.2 & 33.1 & 36.4 \\
DBPedia       & 41.1 & 34.5 & 37.7 & 41.0 & 41.1 & 41.3 & 43.5 & 42.0 & 43.2 & 42.9 & 44.2 \\
NFCorpus      & 31.0 & 26.7 & 29.3 & 33.0 & 33.4 & 31.0 & 32.8 & 32.5 & 33.1 & 33.4 & 33.8 \\
Robust04      & 43.3 & 41.2 & 43.1 & 48.9 & 49.4 & 45.2 & 47.1 & 47.2 & 48.0 & 48.6 & 49.4 \\
SCIDOCS       & 19.2 & 15.1 & 16.7 & 19.7 & 19.8 & 17.7 & 19.0 & 16.4 & 17.8 & 19.1 & 19.9 \\
ArguAna       & 48.0 & 35.6 & 40.6 & 46.2 & 47.6 & 39.8 & 44.0 & 36.4 & 40.5 & 44.8 & 47.0 \\
NQ            & 52.5 & 44.1 & 48.4 & 57.8 & 58.9 & 49.9 & 55.5 & 50.6 & 54.6 & 56.1 & 59.4 \\
BioASQ        & 39.7 & 23.6 & 30.1 & 42.2 & 43.0 & 36.7 & 43.2 & 43.1 & 46.0 & 41.1 & 45.0 \\
Climate-FEVER & 26.8 & 21.6 & 23.6 & 23.3 & 23.5 & 29.9 & 30.8 & 28.7 & 29.4 & 27.7 & 28.7 \\
Signal-1M     & 24.7 & 22.5 & 24.2 & 24.7 & 24.7 & 24.4 & 25.3 & 25.6 & 27.0 & 25.1 & 25.8 \\
\midrule
Avg           & 40.4 & 32.8 & 36.7 & 41.9 & 42.5 & 40.4 & 43.4 & 41.5 & 43.5 & 42.9 & 44.6 \\

\bottomrule
\end{tabular}
\label{tab:beir_supervised_msm}
\end{table*}

\section{Experimental Setup} 

To implement \umbrela, we directly leverage the \textsc{Umbrela} prompt described in ~\citet{upadhyay2024umbrela}.\footnote{We modify the prompt for ArguAna to describe the counter-argument retrieval task.}  Unless otherwise specified, \textsc{Umbrela} is applied to only the top-10 candidate documents retrieved by the initial retriever, following previous work~\cite{lei-etal-2024-corpus, shen-etal-2024-retrieval}; however, we use \emph{at most} eight documents to update the query representation.\footnote{If all candidates are deemed relevant, we simply select the top-8 returned by $R_1$.} This choice matches the number of generated documents from HyDE and \hydeprf, which we describe next. The use of \emph{at most}  eight documents to update the query representation remains fixed across all experiments.\footnote{We say ``at most'' since there will be fewer than eight feedback documents in cases \textsc{Umbrela} finds $< 8$ documents relevant.} 

For HyDE and \hydeprf, we  follow the implementation details outlined by ~\citet{gao2023precise}. We generate eight hypothetical documents using a temperature of 0.7 with a maximum length of 512 tokens. As with \umbrela, unless otherwise specified, we feed \hydeprf the top-10 retrieved documents as in-context exemplars. To handle prompt length limits, we truncate each of the in-context documents to at most 256 tokens. 

As the number of feedback documents for \hybridjudge will always exceed eight, to match the goal of RQ3 while maintaining a fair comparison with the above methods, we prioritize documents identified by \umbrela and use HyDE documents to ``pad'' the difference between eight feedback documents and the number of feedback documents identified by \umbrela. Meaning, for example,  if \umbrela yields five feedback documents, we append three HyDE-generated documents.

We consider three retrieval models across our experiments: BM25, Contriever~\cite{izacard2022unsupervised}, and its supervised fine-tuned variant, Contriever MS-MARCO (Contriever MSM). For each retriever, in addition to the LLM PRF methods, we evaluate traditional PRF baselines which directly utilize the top-8 retrieved documents (to match the number of maximum feedback documents used by all LLM PRF methods) as the set of feedback documents. We also evaluate each retriever without PRF. 

All retrieval experiments were performed in Pyserini~\cite{lin2021pyserini} with default parameters. Qwen3-14B~\cite{yang2025qwen3} served as the backbone LLM and all LLM inference was performed using vLLM~\cite{woosuk2023vllm}.

We evaluate all approaches on 13 low-resource retrieval datasets from BEIR~\cite{thakur2021beir}. The tasks include news retrieval (TREC-News, Robust04), financial question answering (FiQA), entity retrieval (DBpedia), biomedical IR (TREC-Covid, NFCorpus, BioASQ), fact checking (SciFact, Climate-FEVER), citation prediction (SciDocs), tweet retrieval (Signal-1M), question answering (NQ) and argument retrieval (ArguAna). For metrics, we report nDCG@20. 
\section{Varying the Feedback Model}
\label{sec:rq1}

\begin{itemize}
    \item[{\bf RQ1}]   For a given feedback source, how does the choice of feedback model impact PRF effectiveness? 
\end{itemize}

\noindent
To answer the first research question, we evaluate each of the LLM PRF methods (\umbrela, HyDE, \hydeprf, \hybridjudge) across various feedback models. As discussed in Section~\ref{sec:methods}, for sparse retrieval we consider both the Rocchio and RM3 feedback models. We set the Rocchio parameters to $\alpha$=1 and $\beta$=0.75 and for RM3 we set the query weight $\lambda$=0.5.  Lastly, we limit all feedback models to at most 128 feedback terms following \citet{jedidi2025revisiting}. For dense retrieval, we consider both the average vector feedback and Rocchio vector feedback.  For the Rocchio query update, we set $\alpha$=0.4 and $\beta$=0.6 following the parameters used in \citet{li2022pseudo}. The results for this experiment can be found in Table~\ref{tab:beir_unsupervised_bm25_only} for BM25 and Tables ~\ref{tab:beir_unsupervised_contriever}, ~\ref{tab:beir_supervised_msm}, for Contriever and Contriever MSM, respectively. We summarize the key insights below: 

\begin{itemize}[leftmargin=15pt]
    \item For BM25 (Table~\ref{tab:beir_unsupervised_bm25_only}), results show a clear benefit of using the Rocchio feedback algorithm ($\approx$1 point improvement, on average) for HyDE documents versus RM3. This is in contrast to feedback derived from corpus documents, where the two methods are similar with \umbrela (43.3 vs 43.0, for RM3 and Rocchio, respectively), and RM3 is stronger with no LLM involvement (38.7 vs 37.3).
    \item  For dense retrievers (Tables ~\ref{tab:beir_unsupervised_contriever} and  ~\ref{tab:beir_supervised_msm})
    using the Rocchio feedback algorithm improves HyDE's effectiveness when compared to the simple average vector approach utilized by Gao et al.~\cite{gao2023precise}. In particular, HyDE and \hydeprf improve by 4.4 (13.3\%) and 3.1 (9.1\%) points, respectively, when utilizing the Rocchio update instead of the average vector update when applied to Contriever, with similar improvements with Contriever MSM, where HyDE improves by 3 points (7.4\%) and \hydeprf  by 2 points (4.9\%). As before, \umbrela's accuracy remains stable across feedback models.  
\end{itemize}

\noindent
These results demonstrate that the choice of feedback model can be particularly critical to the effectiveness of LLM PRF methods which directly generate feedback terms (i.e., HyDE and \hydeprf), especially in the dense retriever setting. Furthermore, the results  emphasize the importance of evaluating different LLM PRF methods across multiple feedback models, especially when developing new techniques. For example, in the dense retrieval setup, blindly comparing \umbrela with HyDE using average vector feedback may lead one to incorrectly conclude that \umbrela is stronger than HyDE.

As most LLM PRF methods are more effective with the Rocchio feedback model -- and \umbrela performs similar regardless of the feedback model -- subsequent comparisons and experiments will 
focus on the results using Rocchio. 

\section{Corpus-Derived vs. LLM-Generated Feedback}
\label{sec:rq2}

\begin{itemize}
    \item[{\bf RQ2}]  Should feedback be derived from high-quality corpus documents or ``fake'' LLM-generated documents?
\end{itemize}

\noindent
In this section, we answer RQ2 by comparing \umbrela (Feedback: Corpus Only) and HyDE (Feedback: LLM Only) across both sparse and dense retrievers. As the HyDE-generated feedback is ``retriever-free'', it only has access to the downstream retriever ($R_2$). To fairly evaluate \umbrela against HyDE, we begin by restricting \umbrela to a single retriever for both initial and downstream retrieval (i.e., $R_1 == R_2$). We then relax this restriction by allowing \umbrela to use any initial retriever while keeping the downstream retriever fixed to the same $R_2$ as HyDE.

\subsection{Results: Single Retriever Available}
\label{sec:umbrela_singel_ret}

We begin our analysis by comparing both methods when applied for expanding BM25 queries, which is shown in Table ~\ref{tab:beir_unsupervised_bm25_only}.  The results demonstrate that feedback terms derived directly from LLM hypothetical documents are more effective, making a 1.5-point (3.4\%) improvement over directly using high-quality corpus documents. The dense retrieval  setting (Tables~\ref{tab:beir_unsupervised_contriever} and ~\ref{tab:beir_supervised_msm}) shows a similar story: feedback derived from embeddings of LLM-generated hypothetical documents (HyDE) is more effective than feedback derived solely from corpus documents. In particular, with Contriever, HyDE makes a 1.5-point (4.2\%) improvement.  With Contriever-MSM the improvement is smaller, with only a 0.9-point (2.1\%) gain, on average.  

\begin{table}[t!]
  \caption{Bounding the number of HyDE generations to the number of relevant documents found by \umbrela.}
  \centering
  \resizebox{0.75\linewidth}{!}{
  \begin{tabular}{l|c}
    \toprule
    \textbf{Method} & \textbf{BEIR (Avg.)} \\
    \midrule
    \bf Retriever: BM25 \\
    \umbrela & 43.0 \\
    HyDE-Bounded & 43.8 \\
    HyDE-Original & 44.5\\
    \midrule
    \bf Retriever: Contriever \\
    \umbrela & 35.6\\
    HyDE-Bounded & 35.6 \\
    HyDE-Original & 37.1 \\
    \midrule
    \bf Retriever: Contriever-MSM \\
    \umbrela & 42.5\\
    HyDE-Bounded & 42.8\\
    HyDE-Original & 43.4\\
    \bottomrule
  \end{tabular}}
  \label{tab:hyde_ablation}
\end{table}

However, it remains unclear if the differences in effectiveness can be attributed to (1) corpus documents having less useful feedback terms or (2) other factors such as the fact that  \umbrela is bounded by the quality of documents returned by the initial retriever. For example, if BM25 only returns one relevant document in the top-$k$, \umbrela only has one feedback document, whereas HyDE being ``retrieval free'' can always generate a set number  of feedback documents (in our case, eight). In an attempt to control for this difference, we run an ablation in which we \emph{bound} the number of HyDE generations to the number of high-quality documents used by  \umbrela. The results are in Table~\ref{tab:hyde_ablation}. 

\begin{figure}[t]
\centering
\includegraphics[width=0.75\linewidth]{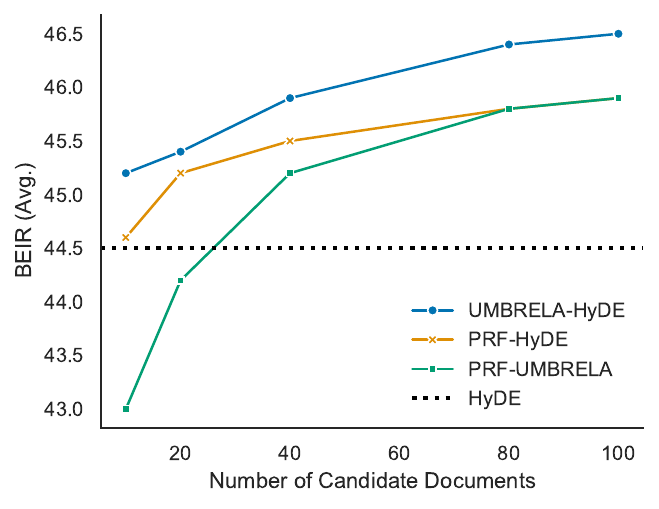}
\caption{Effectiveness of LLM-PRF methods across different number of top-$k$ candidate  documents from BM25. \hybridjudge and \hydeprf are explored in Section~\ref{sec:combining_feedback_tests}.}
\label{fig:judged_docs_vs_acc}
\end{figure}

Under this setup, HyDE and \umbrela are much closer in effectiveness, especially in the dense retrieval setting. This suggests that one explanation for HyDE's improved effectiveness  arises from the fact that it has more sources to extract feedback terms (or vectors) from, not that it necessarily has better feedback content. To further examine this hypothesis, we investigate a simple solution to enable \umbrela access to more feedback sources: apply \umbrela to judge more candidate documents from the initial retriever. However, we note that we \emph{still} limit the number of feedback documents used for updating the query to at most eight documents regardless of the number of candidate documents judged, matching the number of HyDE-generated documents. We focus on BM25 as the retriever for this experiment as it consistently demonstrated the strongest PRF effectiveness across methods. The results of this experiment can be found in Figure~\ref{fig:judged_docs_vs_acc}. 

The results show a consistent gain in \umbrela's effectiveness as the number of candidate documents increases. At around 20 assessed candidate documents, \umbrela becomes on par with HyDE. After 40 assessed candidate documents, \umbrela begins to surpass HyDE. While simply judging \emph{more} documents is a solution, a natural question arises regarding its efficiency. We explore this in more detail in Section \ref{sec:efficiency}.

\subsection{Results: \umbrela with a Stronger Initial Retriever}

\label{sec:umbrela_strong_ret}

\begin{table}[t!]
\caption{The impact of the initial retriever, $R_1$, on the effectiveness of \umbrela, \hydeprf, and \hybridjudge. HyDE as $R_1$ refers to HyDE with BM25.}
\centering
\resizebox{0.75\linewidth}{!}{
\begin{tabular}{cl|c}
\toprule
 & \textbf{Method} & \textbf{BEIR (Avg.)} \\
\midrule
\multicolumn{3}{c}{\textbf{$R_1$: N/A $\rightarrow$ $R_2$: BM25}} \\
\midrule
1 & BM25 & 36.5 \\
2 & HyDE & 44.5 \\
\midrule
\multicolumn{3}{c}{\textbf{$R_1$: BM25 $\rightarrow$ $R_2$: BM25}} \\
\midrule
3 & \umbrela & 43.0 \\
4 & \hydeprf & 44.6 \\
5 & \hybridjudge & 45.2 \\
\midrule
\multicolumn{3}{c}{\textbf{$R_1$: HyDE $\rightarrow$ $R_2$: BM25}} \\
\midrule
6 & \umbrela & 46.1 \\
7 & \hydeprf & 46.1 \\
8 & \hybridjudge & 46.2 \\
\midrule
\midrule
\multicolumn{3}{c}{\textbf{$R_1$: N/A $\rightarrow$ $R_2$: Contriever MSM}} \\
\midrule
9 & Contriever MSM & 40.4 \\
10 & HyDE & 43.4 \\
\midrule
\multicolumn{3}{c}{\textbf{$R_1$: Contriever MSM $\rightarrow$ $R_2$: Contriever MSM}} \\
\midrule
11 & \umbrela & 42.5 \\
12 & \hydeprf & 43.5 \\
13 & \hybridjudge & 44.6 \\
\midrule
\multicolumn{3}{c}{\textbf{$R_1$: Contriever MSM $\rightarrow$ $R_2$: BM25}} \\
\midrule
14 & \umbrela & 46.1 \\
15 & \hydeprf & 47.3 \\
16 & \hybridjudge & 47.2 \\
\bottomrule
\end{tabular}}
\label{tab:initial_ret_beir_avg}
\end{table}

In the previous subsection, to fairly evaluate \umbrela and HyDE, we restricted \umbrela's initial retriever to be equivalent to HyDE's downstream retriever. However, in practical settings \umbrela  is not bounded by such limitations.  In this subsection, we go beyond this ``fair'' setup to understand how leveraging feedback from high-quality corpus documents compares to HyDE when using an initial retriever which generates stronger candidate documents. We note that this setup was somewhat explored with Contriever MSM. However, given that PRF with BM25 was more effective than with Contriever MSM, it is still unclear if the explanation for HyDE's improvement over \umbrela is due to better candidate documents not being more helpful for PRF or simply because dense retrievers are less capable of taking advantage of PRF. To address this,  we focus on BM25 as the downstream retriever,  while varying the  retriever which produces the candidate documents. 

The results are shown in Table~\ref{tab:initial_ret_beir_avg}. Overall, we see an expected trend: the stronger candidate documents returned by HyDE with BM25 and Contriever MSM consistently improve the effectiveness of \umbrela with BM25. Beyond this general trend, we make the following observations: 

\begin{itemize}[leftmargin=15pt]

\item  The difference in effectiveness of \umbrela with HyDE as $R_1$ (row 6) and HyDE (row 2) suggests that feedback terms derived from corpus documents retrieved by HyDE are more useful for PRF than feedback terms generated by HyDE.  
\item Feedback documents from Contriever MSM are better for PRF, but Contriever MSM's query encoder is less capable of exploiting this feedback as compared to BM25's query encoder. Particularly, when refining the query based on feedback from Contriever MSM's top returned documents, BM25 outperforms Contriever MSM by 3.6 points (8.5\%) with \umbrela (row 14) compared to using \emph{the same feedback} to update   Contriever MSM's query (row 11). Notably, this improvement is in spite of the fact that BM25 additionally starts off 3.9 points (10\%) worse than Contriever MSM. We note that a similar trend is observed with HyDE, which,  being ``retrieval free'', utilizes the same hypothetical documents regardless of whether it's updating the query for BM25 or Contriever MSM. Interestingly, however, the difference between HyDE with BM25 and HyDE with Contriever MSM is much smaller with a 2.5\% relative difference (44.5 versus 43.4 nDCG@20).

\item The results of utilizing Contriever MSM  to improve queries for BM25 highlights an interesting insight: PRF in the dense vector space may be suboptimal and feedback from dense retrievers may be more effectively exploited by a sparse retriever like BM25. How this finding extends to other retrievers, especially stronger dense models, remains an open question. We leave this to future work. 
\end{itemize}

\noindent
Taken together, the results from Section~\ref{sec:umbrela_singel_ret} and  ~\ref{sec:umbrela_strong_ret} provide a nuanced answer to RQ2. Setting $R_2$ to BM25 as the representative case, our findings show the following. If restricted to a single retriever, feedback from the corpus is only more or equally as effective as HyDE if scaling the number of candidate documents judged by \textsc{Umbrela}. As we will show in Section~\ref{sec:efficiency}, this comes at a much higher latency. On the other hand, improving $R_1$ presents an alternative solution where \umbrela can outperform HyDE. Thus, to answer RQ2 concretely, while feedback derived from high-quality corpus documents \emph{can} be more effective, it is heavily reliant on either a strong initial retriever or increased inference-time (i.e., ``query-time'') latency. In contrast, LLM-generated hypothetical answer documents offer a simpler and practical solution that better balances effectiveness and efficiency.   
\begin{figure*}[t!]
  \centering
  \includegraphics[width=0.8\textwidth]{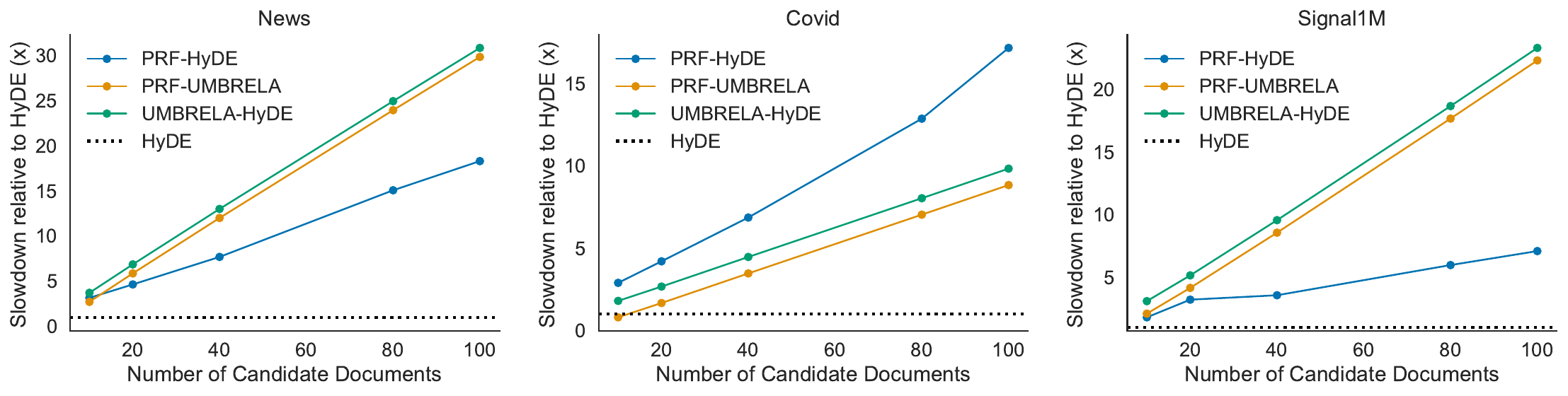}
  \caption{Latency of LLM PRF methods relative to HyDE.}
  \label{fig:latency}
\end{figure*}

\section{Combining Feedback from Corpus \& LLM}
\label{sec:combining_feedback_tests}
\begin{itemize}
    \item[{\bf RQ3}] Does combining feedback sources improve PRF effectiveness over using a single source? If so, how should the feedback be combined?
\end{itemize}

\noindent
In Section \ref{sec:rq2} we examined how feedback derived solely from the corpus compares to feedback generated solely from the LLM's parametric knowledge. We now consider the combination of the two feedback sources, with a focus on first understanding  the effectiveness ``boost'' of incorporating the additional feedback source to \umbrela and HyDE and then also the optimal strategy for combining feedback. To do so, in this section, the primary  comparisons are between (1) \umbrela and \hybridjudge; (2) HyDE and \hydeprf; and lastly, (3)  \hybridjudge and \hydeprf, two approaches for combining the feedback sources. At a high-level, \hybridjudge augments the feedback from \umbrela by \emph{explicitly} concatenating vectors independently derived by \umbrela and HyDE. \hydeprf incorporates corpus evidence \emph{implicitly} by prompting HyDE with corpus documents as in-context examples. 

\subsection{Results: Single Retriever Available}
\label{sec:hyde-prf_umbrela_hyde_single_ret}

As in Section ~\ref{sec:umbrela_singel_ret}, we begin in the setting in which a single retriever is available. Starting with Table~\ref{tab:beir_unsupervised_bm25_only},  we first note that surprisingly, there is no improvement when prompting HyDE with in-context examples from BM25, as both HyDE and \hydeprf are equally effective on average (44.5 versus 44.6 nDCG@20).  On the other hand, combining PRF signals from the corpus and LLM via the \hybridjudge method makes a 2.2-point (5\%) improvement on \umbrela.  However, the improvement of \hybridjudge upon \hydeprf is minimal, with only a 0.6-point (1.3\%) improvement. 

With dense retrievers (Tables 2 and 3), we again see no improvement when prompting HyDE with in-context examples. However, we do find that \hybridjudge makes a strong improvement in effectiveness versus \umbrela, HyDE and \hydeprf.  For example, \hybridjudge makes a 2.9-point (7.8\%) and 1.1-point (2.5\%) improvement upon \hydeprf with Contriever and Contriever MSM, respectively.  

How does integrating more candidate corpus documents impact accuracy?   In Figure~\ref{fig:judged_docs_vs_acc}, the results demonstrate that both \hydeprf and \hybridjudge  improve with more candidate documents, with \hydeprf seeing a larger initial benefit. However, the differences in effectiveness between  \hydeprf and \hybridjudge versus \umbrela decrease as more candidate documents are scored, with all methods being about equally effective (46.5, 45.9, and 45.9 nDCG@20,  respectively). 

Interestingly, it is of note that for methods like \hydeprf, it has been shown that there is marginal improvement in accuracy when the number of candidate examples exceeds 10~\cite{shen-etal-2024-retrieval}. Our results seem to show the opposite:  \hydeprf demonstrates improvements up until about $\approx 40$ candidate documents. This is likely due to improvements in the ability for LLMs to handle longer contexts. 

\subsection{Results: \hybridjudge and \hydeprf with a Stronger Initial Retriever}
\label{sec:hyde-prf_umbrela_hyde_multi_ret}
We next explore how \hybridjudge and \hydeprf benefit from  improving the initial retriever. As before, the results are in Table ~\ref{tab:initial_ret_beir_avg}. We summarize the key insights below: 

\begin{itemize}[leftmargin=15pt]
    \item Comparing \hybridjudge (row 5) with \umbrela (row 6) suggests that if combining feedback from the LLM and corpus, there is a modest improvement (2\%) by doing so in a two-step approach (i.e., HyDE then \umbrela), rather than combining each source independently. 
    \item Interestingly, under this two-step approach there is no further improvement when ``padding'' \umbrela with HyDE vectors, as \umbrela (row 6) and \hybridjudge (row 8) are equally effective (46.2 versus 46.1).  We hypothesize that this may be due to the redundancy of terms present in the HyDE-retrieved documents. High scoring documents retrieved by HyDE with BM25 will likely contain similar terms to those in the HyDE-generated documents due to the exact-matching scoring scheme.
    \item The results in Section \ref{sec:hyde-prf_umbrela_hyde_single_ret} suggested that there is no advantage to incorporating in-context examples for HyDE. However, we find that when using stronger in-context examples, \hydeprf is more effective than HyDE. Using documents retrieved by HyDE with BM25 as input to \hydeprf (row 7) yields a 1.6-point (3.6\%) improvement upon HyDE (row 2) and using documents retrieved by Contriever MSM (row 15) yields even more gains, with a 2.8-point (6.3\%) improvement. 
    \item It is of note that \hydeprf seems to yield further gains using documents retrieved by Contriever-MSM versus documents retrieved by HyDE despite HyDE having higher retrieval accuracy. We hypothesize that the reason lies in the fact that the documents retrieved by HyDE with BM25 are likely less \emph{diverse} (i.e., more redundant), being biased towards real documents that ``look like'' those sampled by HyDE, which tend to generally cover similar content across generations. This result shows that better retrieval does not \emph{necessarily} mean better \hydeprf effectiveness and other factors may be important.
\end{itemize}

\noindent
To answer RQ3, we found the answer to depend on the downstream retriever. For dense retrieval setups, we found that \hybridjudge demonstrated consistent improvement over methods which utilized a single source of feedback, providing a clear answer: for dense retrieval, combining feedback sources is helpful and it should be done by independently combining feedback derived from the corpus and the LLM. 

With BM25, the answer is not as obvious. We found the combination of feedback sources to be most helpful under two cases. The first is if the feedback is combined \emph{sequentially} (i.e., run HyDE and then \umbrela) and not in parallel. The second case is if a strong initial retriever (without any PRF) generates the set of candidate corpus documents. Under each of the two cases, we did not see any strong differences in effectiveness based on \emph{how} the feedback was combined, meaning the choice of combination approach should be based on other practical considerations, such as latency, which is discussed in the next section. Interestingly, we also note that these results suggest that,  for PRF with BM25, strong corpus documents are necessary to add additional relevance feedback signal beyond the feedback already derived from the LLM's parametric knowledge.

\section{Latency Comparisons}
\label{sec:efficiency}

\begin{itemize}
  \item[{\bf RQ4}]  How does the latency vary among LLM PRF methods?
\end{itemize}

\noindent
A practical consideration when utilizing LLMs for PRF is the efficiency.  In this section, we compare the latency of each of the LLM PRF methods relative to HyDE. We primarily focus on three BEIR datasets which span a range of document lengths. For documents on the longer side, we use TREC-News, which has about 634.8 words per document; TREC-Covid, which has 160.8 words per document, represents the average length ($\approx$ 200 words per document across the 13 BEIR datasets evaluated); and lastly, for shorter documents, we use  Signal-1M, which has about 13.9 words per document~\cite{thakur2021beir}. For this experiment, we only compute the latency of the LLM component; thus the slowdown of each method is measured based on the ratio of its latency to that of HyDE. For \hybridjudge, we consider its ``worst-case'' latency, assuming that both \umbrela and HyDE are run one after the other, and not in parallel. Experiments are run on NVIDIA RTX 6000 Ada Generation GPUs using vLLM~\cite{woosuk2023vllm}. 

The results can be found in Figure~\ref{fig:latency}. Overall, the results show that each of the approaches which leverage corpus knowledge are generally less efficient than HyDE. This slowdown increases as more candidate documents are considered, meaning that while it was demonstrated in Figure~\ref{fig:judged_docs_vs_acc} that more candidate documents can increase accuracy, this comes at a much slower latency with respect to HyDE. 

Comparing \umbrela, \hydeprf, and \hybridjudge, the answer appears to depend on the document lengths in the corpora.   For corpora which contains long documents (TREC News), the three approaches are comparable at a lower number of candidate documents (10), but \umbrela (and with it,  \hybridjudge) begins to be much slower as the number of candidate documents it assesses increases. Corpora with shorter documents (Signal1M) shows a similar trend: \umbrela and \hydeprf  begin at a similar latency, but \umbrela is slower at higher numbers of candidate documents. On datasets with average length documents (Covid), \umbrela and \hybridjudge are consistently faster than \hydeprf across all amounts of candidate documents. 

\section{Conclusion}

In this study, we conduct a large-scale and controlled evaluation of pseudo-relevance feedback with LLMs, disentangling the independent roles that the \emph{feedback source} and the \emph{feedback model} --  two central ingredients of modern PRF methods -- play. Our results demonstrate that each of these dimensions influences the downstream effectiveness of LLM PRF methods. In particular, we found that the effectiveness of methods which leverage feedback generated directly by an LLM (i.e., HyDE or \hydeprf), can be dependent on the choice of feedback model, especially in dense retriever settings. If utilizing feedback derived solely from -- or in combination with -- corpus documents, the most important factor is the effectiveness of the initial retriever used to obtain candidate documents. Such methods are most effective when candidate documents are retrieved by a strong initial retriever, especially when accounting for practical factors like latency.  Overall, our results provide a \emph{fair} comparison between various LLM PRF approaches under a unified experimental framework, providing insights for practitioners and  guiding best practices for future LLM-based PRF approaches.  

\section*{Acknowledgments}

This research was supported in part by the Natural Sciences and Engineering Research Council
(NSERC) of Canada. We would like to thank Felix Labelle and Daniel Gwon for their thoughtful feedback and contributions. 
\bibliographystyle{ACM-Reference-Format}
\balance
\bibliography{main}
\appendix

\end{document}